\documentclass[a4paper]{article}

\usepackage{INTERSPEECH2019}
\usepackage{dsfont}
\usepackage{graphicx}
\usepackage{xcolor}
\usepackage{caption}
\usepackage{subcaption}
\usepackage{multirow}
\usepackage{cite}
\usepackage{color,soul}

\title{Identifying Mood Episodes Using Dialogue Features from Clinical Interviews}

\name{Zakaria Aldeneh\(^{1}\), Mimansa Jaiswal\(^{1}\), Michael Picheny\(^3\),\\
Melvin McInnis\(^{2}\),Emily Mower Provost\(^{1}\)}
\address{Departments of: \(^1\)Computer Science and Engineering and \(^2\)Psychiatry, 
University of Michigan
\\ \(^3\)IBM T. J. Watson Research Center}
\email{\texttt{\{aldeneh, mimansa, mmcinnis, emilykmp\}@umich.edu,
        picheny@us.ibm.com}}
\begin{document}
\maketitle


\begin{abstract}
Bipolar disorder, a severe chronic mental illness characterized by pathological mood swings from depression to mania, requires ongoing symptom severity tracking to both guide and measure treatments that are critical for maintaining long-term health. Mental health professionals assess symptom severity through semi-structured clinical interviews.  During these interviews, they observe their patients' spoken behaviors, including both what the patients say and how they say it. In this work, we move beyond acoustic and lexical information, investigating how higher-level interactive patterns also change during mood episodes. We then perform a secondary analysis, asking if these interactive patterns, measured through dialogue features, can be used in conjunction with acoustic features to automatically recognize mood episodes. Our results show that it is beneficial to consider dialogue features when analyzing and building automated systems for predicting and monitoring mood.

\end{abstract}


\section{Introduction}
Bipolar disorder (BP) is a mental disorder that affects more than 20 million people 
worldwide~\cite{kessler2005lifetime}. 
Individuals who suffer from BP experience mood episodes that range from mania (elevated mood) 
to depression (lowered mood).
These mood episodes have negative consequences on an individual's life and work. 
The current standard for monitoring symptom severity is through regular clinical interviews.
In these interviews, a mental health professional asks a series of questions that address 
different aspects of a patient's life and mood.
The success of these interviews relies on how a clinician interprets both the verbal 
and non-verbal cues of a patient's response to the open-ended interview questions, 
making the interviews subjective in nature.
In addition, regular clinical appointments can be both costly and time consuming.
Recent advances in affective computing have made it possible to develop automated methods 
to help make the current standard not only more objective, 
but also more accessible to a wider range of people. Automated approaches can also help 
quantify patient's behavior and provide clinicians with actionable data.
This work focuses on leveraging interaction patterns from clinical interviews 
for automatically recognizing mood episodes.

Previous research showed that patient mood episodes affect the acoustics of both the patients and
the clinicians in clinical interviews.  For instance, variations in pitch,
speaking rate, and intensity in a patient's speech were shown to correlate 
with symptom severity 
\cite{yang2013detecting, yu2013multimodal, al2018detecting, faurholt2016voice, khorram2016recognition, karam2014ecologically, gideon2016mood}.
Given the dyadic nature of the clinical interviews, 
changes in speech patterns of the patient affect 
speech patterns of the clinician through the phenomenon of entrainment
\cite{levitan2011measuring, pardo2006phonetic, niederhoffer2002linguistic,nasir2018towards,xiao2015analyzing}. 
These observed changes in speech patterns of both the patient and the clinician in the dialogues affect the dynamics of the interaction~\cite{hall1995nonverbal}. 
In other words, we expect patients' mood episodes to affect not only patient and clinician acoustics, but also the turn-taking behaviour in clinical dialogues.
The goal of this work is to study the effect of mood changes on interaction dynamics in clinical interviews. We show that the extracted high-level dialogue features can be used to augment prosodic features to improve the performance of automatically detecting depression severity in clinical interviews. 
Our results show that mood measurably affects interaction patterns, and that it can be beneficial to
take dialogue features into account when building automated agents for conducting clinical interviews.


\section{Related Work}

Although the use of acoustic bio-markers has not yet been embraced in clinical practice, there has been extensive research done on the topic.
We refer the reader to a paper by Cummins et al.~\cite{cummins2015review} for a 
more comprehensive literature review.

\textbf{Prosody.} Prosodic features capture variations in rhythm, stress, and 
intonation of speech. They represent information present in syllables and
larger units of speech (i.e., not at the phone level). 
A study by Hashim et al.~\cite{hashim2017evaluation} showed that acoustic measures that capture timing properties of speech are predictive of depression scores. Specifically, the authors showed that
features that capture transitions between different voicing properties can be effective 
for predicting clinical depression scores. Many studies have shown the 
effectiveness of  pitch and rhythm features for detecting mood state
~\cite{yang2013detecting,karam2014ecologically,scherer2014dyadic,
gideon2016mood,faurholt2016voice}.

\textbf{Interaction.} 
Modulations in patients' speech patterns affect clinicians' speech 
patterns during clinical interviews~\cite{yang2013detecting}.
Acoustic features extracted from the clinician during clinical
interviews have been shown to correlate with the mood symptom severity of patients.
For instance, Scherer et al.~\cite{scherer2014dyadic} showed that acoustic 
features extracted from interviewers varied with patients' depression severity.
The authors also found that some entrainment measures had an inverse 
relationship with depression severity. 
Dibeklioglu et al.~\cite{dibekliouglu2015multimodal} and Yu et al.~\cite{yu2013multimodal}
showed that conversational features (e.g., onset time and utterance durations)
can be effective for detecting mood symptom severity in patients.

In contrast to previous work, the novelty of our work is three-fold: (1) we introduce a set
of dialogue features to aid in the prediction of mood symptom severity; 
(2) we analyze dialogue features using a linear mixed effect model to 
study how mood episodes affect interaction patterns; 
(3) we show that explicitly adding high-level dialogue features to acoustic-based systems
can improve the performance of automatic mood symptom severity prediction.



\section{The Data}
The PRIORI (Predicting Individual Outcomes for Rapid Intervention) dataset is
a longitudinal dataset composed of cellphone recordings collected as part of a large-scale effort to 
study how the properties of speech change with mood symptom severity~\cite{karam2014ecologically}.
The participants in the study include individuals who are diagnosed with type-I or type-II BP. 
Participants were provided with a smartphone equipped with specialized secure recording software that they used as their
primary cellphone during the duration of the study (maximum duration of 13-months). 
The software records only their side of the conversation for both incoming and outgoing calls.  
It encrypts the speech in real-time and then uploads the data to a HIPAA compliant server for offline processing.

The data include two types of calls: assessment calls and personal calls. 
Assessment calls are weekly calls between a participant and a 
clinician in which the clinician evaluates the participant's mood
using the the Hamilton Depression Scale (HAMD)~\cite{hamilton1967development} 
and the Young Mania Rating Scale (YMRS)~\cite{young1978rating} 
as part of a clinical interview.
YMRS is a rating scale used for assessing mania severity while HAMD is a rating
scale used for assessing depression severity. In our dataset, both YMRS and HAMD scores 
range from 0 (least symptomatic) to 35 (most symptomatic).
Whereas personal calls were only collected from patient-side cellphone microphones, 
assessment calls were collected from two sources: 
patient-side cellphone microphones, and clinician-side landline telephone recorder microphone.
This gives us the data needed for studying clinician-patient interactions.

The assessment call component of the PRIORI dataset contains over 380 hours of speech from over 1,280 
assessment calls.
Following the work of Gideon et al.\cite{gideon2016mood}, 
we define three mood episodes based on the HAMD and YMRS scores:
euthymic state (YMRS~$\le$~6 \& HAMD~$\le$~6),
depressed state (YMRS~$\le$~6 \& HAMD~$\ge$~10),
and manic state (YMRS~$\ge$~10 \& HAMD~$\le$~6).
We exclude all calls outside of these ranges.
Our final dataset includes 155 hours of 
speech from 641 calls (317 euthymic, 268 depressed, 56 manic) by 47 unique 
speakers (34 females) and 9 unique clinicians (5 females).

\subsection{Extracting Speaker Turns}
The patient-side cellphone recordings of the assessment calls contain single channel streams
with the patient's speech signal. 
The clinician-side landline telephone recordings of the assessment calls contain single 
channel streams with both the clinician's and the patient's speech signals.
The goal is to obtain the start and end times of each speech segment in the conversation.
To do so, we cross-correlate the two signals in the frequency domain 
and use the maximum value as the alignment offset. Once the two signals are aligned,
we run COMBO-SAD, a voice activity detection (VAD) algorithm by Sadjadi and Hansen
\cite{sadjadi2013unsupervised}, to extract speech segments from the two aligned signals.
The VAD output from the cellphone recordings gives the patient turns, while
the VAD output from the landline recordings gives the merged patient and
clinician turns. Regions where the two VAD signals overlapped were assigned
``patient speech''. Regions where the two VAD signals did not overlap were
assigned ``clinician speech''. 
To form speaker turns, we merge speech activity from a single speaker whenever there is
silence that is \textless500 milliseconds separating speech segments. 
We pick this value as conventional spoken dialog systems use a silence 
threshold of around 500 milliseconds to determine utterance end-points
\cite{arsikere2015enhanced}.


\begin{table*}[ht]
\caption{List of investigated dialogue features with coefficients and standard errors from LMEMs. 
The main effect coefficients indicate changes from euthymia to depression (or from euthymia to mania).
We report $p$-values obtained from likelihood ratio test against a null model with no mood effect.
\textbf{Bolded} estimates indicate significance after correcting the FDR at \(\alpha=0.05\)
\cite{benjamini1995controlling}.
We use `--' to denote estimates for features that showed obvious deviations from homoscedasticity or normality
after visually inspecting the residual plots.
Unless noted otherwise, all values for time-based features are reported in milliseconds. All ratios are reported 
as percentages (\(\%\)).\\
$p$-value codes: `***' $<$0.001; `**' $<$0.01; `*' $<$0.05; `.' $<$0.1}
    \centering
    \label{tab:analysis_results}
    \begin{tabular}{@{}lccccc@{}}
        \toprule
                 & \multicolumn{2}{c}{\textbf{Depression}} &  & \multicolumn{2}{c}{\textbf{Mania}}  \\
        \textbf{Feature} & \textbf{Estimate} & \textbf{$p$-value} &  & \textbf{Estimate} & \textbf{$p$-value} \\
        \midrule
        call duration in minutes$^\dagger$ & \textbf{0.578$\pm$0.052} & *** & & \textbf{0.468$\pm$0.073} & N/A  \\
        number of turn-switches per min. & -0.567$\pm$0.319 & . & & 0.004$\pm$0.526 &   \\

        \textit{patient features} \\
        floor control ratio & \textbf{2.657$\pm$1.173} & * & & \textbf{8.276$\pm$1.981} & *** \\
        hold offset (mean) & -0.263$\pm$77.601 &   & & -215.175$\pm$143.812 &   \\
        hold offset (SD) & -- &   & & -- &   \\
        number of continuous turns (mean) & \textbf{0.072$\pm$0.020} & *** & & 0.066$\pm$0.034 & . \\
        number of continuous turns (SD) & \textbf{0.104$\pm$0.041} & * & & 0.140$\pm$0.067 & * \\
        switch offset (mean) & 8.434$\pm$26.940 &   & & -48.812$\pm$41.987 &   \\
        switch offset (SD) & \textbf{63.812$\pm$20.925} &  ** & & 49.700$\pm$34.688 &   \\
        turn lengths (mean) & 75.175$\pm$52.825 &   & & \textbf{313.163$\pm$84.050} & *** \\
        turn lengths (SD) & 12.875$\pm$73.000 &   & & \textbf{299.200$\pm$117.013} & * \\        

        \textit{clinician features} \\
        hold offset (mean) & \textbf{128.150$\pm$30.200} & *** & & -16.799$\pm$44.070 &   \\
        hold offset (SD) & \textbf{200.262$\pm$39.575} & *** & & -33.312$\pm$55.038 &   \\
        number of continuous turns (mean) & \textbf{0.054$\pm$0.022} & * & & -0.027$\pm$0.034 &   \\
        number of continuous turns (SD) & 0.052$\pm$0.033 &   & & -0.039$\pm$0.052 &   \\
        switch offset (mean) & 6.474$\pm$33.411 &  & & -90.675$\pm$50.237 & . \\
        switch offset (SD) & 55.075$\pm$81.487 &  &  & -- &   \\
        turn lengths (mean) & \textbf{-97.525$\pm$41.600} & *  & & \textbf{-215.300$\pm$70.100} & ** \\
        turn lengths (SD) & -20.137$\pm$38.688 &   & & \textbf{-175.863$\pm$61.812} & ** \\
        \bottomrule
    \end{tabular}\\
    $^\dagger$ Produced significant (\(p<0.0005\)) interaction effect with patient gender control variable in manic episodes.\\
    The estimate for males in the manic state was $+$0.631$\pm$0.161 minutes.
\end{table*}

\section{Features}

\subsection{Dialogue Features}
We extract a set of high-level dialogue features to quantify the 
patient-clinician interactions motivated by the fact that patients experiencing different mood episodes
display less expressive interactive styles, increased talkativeness, 
racing thoughts, and inflated self-esteem~\cite{hall1995nonverbal, cassidy1998signs}.
All of the dialogue features that we study in this work are
time-based features and can be easily extracted using a conventional VAD\@.
This makes the extracted dialogue features more robust to noisy conditions in the recordings when compared 
to features that are extracted directly from the acoustic signal.

\noindent\textbf{Floor control ratio.}
This feature measures the relative amount of time an individual spends speaking to the total 
amount of speech in a conversation. 
Floor control has been studied in the entrainment, turn-taking, and dialogue literature
\cite{meshorer2016using, bevnuvs2011pragmatic}.
This feature can quantify dominance, brevity, and relative duration of the patient's 
response to the interview questions.

\noindent\textbf{Turn hold offset.}
This feature measures the duration of pauses that are less than half a second within turns from the same speaker.
Turn hold offset is a well-studied feature in the turn-taking literature
\cite{levinson2015timing, heldner2010pauses}.
Previous work showed that depressed individuals tend to have longer pauses in their
speech~\cite{hashim2017evaluation}.

\noindent\textbf{Number of consecutive turns.}
This feature measures the tendency of a speaker to hold the floor in a conversation.
In other words, this feature measures the tendency for a speaker to include
long pauses ($>$500 milliseconds) between his or her sentences.

\noindent\textbf{Number of turn switches per minute.}
The current data segmentation approach makes measuring durations (or frequencies) of overlapping speech difficult.
Previous research, however, showed that the number of turn switches is correlated with the number of interrupts and overlaps in a 
conversation~\cite{shriberg2001observations, yella2014overlapping}.
We use this feature as a proxy for the amount of overlapping speech that occurs in a clinical
interview.

\noindent\textbf{Turn switch offsets.}
This feature measures the latency between turn transitions. Previous work
showed that different dialogue contexts have different turn switch latencies
\cite{heeman2017turn}. Clinically, previous work demonstrated that depressed 
individuals take longer to respond to clinicians' 
questions~\cite{yu2013multimodal}.

\noindent\textbf{Turn lengths.}
This feature measures the duration of each turn by a speaker. 
Previous research showed that variants of this feature were effective for
detecting depression~\cite{yu2013multimodal}.

We summarize dialogue features by taking the mean and standard deviation across 
each conversation. 
This results in 20 features representing the interactions in each clinical interview.

\subsection{Rhythm Features}
Previous work showed that speech rhythm features are effective for predicting
mood states~\cite{gideon2016mood, yang2013detecting, khorram2016recognition}.
We follow the approach mentioned in~\cite{gideon2016mood}
and extract seven rhythm features using the algorithm proposed by 
Tilsen and Arvaniti~\cite{tilsen2013speech}. 
These rhythm features capture power distribution, rate, and rhythm stability metrics. 
To obtain call-level features, we calculate statistics over the seven rhythm features, including: 
mean, standard deviation, kurtosis, skewness, max, min and their normalized  locations,
linear regression slope, intercept, and error.  
This results in a total of 70 features representing each assessment call.


\section{Analyzing the Dialogue Features}
We run a series of linear mixed effects models (LMEMs), using the \textit{lme4}~\cite{bates2007lme4} 
package in R~\cite{r_library}, to analyze the effect of mood on turn-taking in clinical dialogues. 
We only consider clinical interviews with female clinicians in this analysis (\texttildelow$82\%$ of 
the interviews). This obviates the need for considering three-way 
interactions between the predictors (i.e., $mood\times gender_{patient} \times gender_{clinician}$). 
We set each of the dialogue features as a response variable in our linear models. 
Depending on the task, we set the binary mood state 
$\{euthymic, depressed\}$ or $\{euthymic, manic\}$ as a fixed effect test variable. 
We set the gender of the patient as a fixed effect  control variable. 
Finally, we set random intercepts for the patients and the 
clinicians.
We use likelihood ratio tests to test for statistical significance. For each dialogue feature,
we test a full model (with the mood fixed effect) against a null model (without the mood fixed effect).
In the case of a significant interaction between mood state and patient gender, 
we report the $p$-values for the interaction effect since those from the main effect
are not interpretable~\cite{zar1999biostatistical}.

Our final data that we use for this analysis contain 525 clinical interviews with
a total duration of around 130 hours from 46 unique patients (33 females) and 5 unique clinicians
(all females).

\subsection{Results and Discussion}
We report the results of the LMEM analysis in Table~\ref{tab:analysis_results}.

\subsubsection{Depression}
We find that call duration goes up by an average of
\(0.578\pm0.052\) minutes when patients are in a depressed episode 
(compared to a ehuthymic episode).
Call duration did not produce a significant 
interaction effects with the gender effect, showing that the increase in 
duration is consistent across male and female patients.

\textbf{Patient Features.}
Floor control of patients significantly goes up by \(2.657\%\pm1.173\%\)
when patients are depressed.
We also find that both the mean and variability of the number of continuous turns
go up by an average of \(0.072\pm0.020\) and  \(0.104\pm0.041\) respectively.
Additionally, the variability in turn switch offset for the patients 
goes up by \(63.812\pm20.925\) milliseconds when patients are depressed.
None of the patient features demonstrated statistically significant interactions with gender.
The results suggest that depressed patients are more likely to insert longer pauses
($>$500 milliseconds) while speaking. Additionally, the results suggest that
depressed patients exhibit higher variability in the time they take to respond to 
questions by clinicians.

\textbf{Clinician Features.}
We find that both the mean and variability of turn hold offsets in clinicians
go up by \(128.150\pm30.200\) milliseconds and \(200.262\pm39.575\) milliseconds,
respectively. We find that the number of continuous turns for clinicians goes up 
by an average of \(0.054\pm0.022\).
Finally, we find that the average turn length
of clinicians goes down by \(97.525\pm41.600\) milliseconds.
There were no significant interaction effects.
These results suggest that although clinicians insert longer silences between their turns
while interviewing depressed patients, they tend to speak for a slightly
shorter time.

\subsubsection{Mania}

We find that call duration goes up by an average of \(0.468\pm0.073\) minutes
when patients are in a manic episode (compared to a ehuthymic episode).
Call duration produced significant interaction with gender
(\(p<0.0005\)), indicating that the increase in call duration is mainly driven 
by male patients (\(+0.631\pm0.161\) minutes).

\textbf{Patient Features.} We find that floor control of patients significantly 
goes up by an average of \(8.276\%\pm1.981\%\) when patients are manic.
We find that both the mean and variability of turn lengths in patients
go up by \(313.163\pm84.050\) milliseconds and \(299.200\pm117.013\) milliseconds,
respectively.
None of the patient features produced significant interactions with their gender, 
meaning that both male and female patients speak longer relative to clinicians
when patients are manic.

\textbf{Clinician Features.}
We find that both the mean and variation in clinicians' turn lengths go down by
\(215.300\pm70.100\) and \(175.863\pm61.812\) milliseconds, respectively. 
This finding was consistent for clinicians interviewing both male and female patients.


\section{Predicting Mood Episodes}
The previous section demonstrated that symptom severity has a significant impact on the dynamics of interaction as captured by our dialogue features.
In this section, we assess whether mood symptom severity prediction can be improved by integrating dialogue features with standard prosodic features. 
We define two classification tasks, similar to~\cite{gideon2016mood}: (1) discriminate between episodes of euthymia and depression and (2) discriminate between episodes of euthymia and mania.
We only include patients in this analysis if they have at least two euthymic calls and 
two manic/depressed assessment calls.

We study the efficacy of dialogue features using three classifiers:
logistic regression, support vector machines (SVM), and deep neural networks (DNN).
We train each classifier with rhythm features, dialogue features, and their combination
via early fusion.
We follow a leave-one-speaker-out evaluation scheme, and report the average
area under the receiver operating characteristic curve (AUROC) across all test speakers.
For each test speaker, we run a five-fold cross-validation over the training speakers to pick 
optimal hyper-parameters. 
We build and train our classifiers using the Scikit-learn library~\cite{pedregosa2011scikit}.
We optimize for the following hyper-parameters:
Logistic regression: \{\(C\): [0.001, 0.01, $\dots$, 1000]\};
SVM: \{\(kernel\): [\(rbf\)], \(C\): [0.001, 0.01, $\dots$, 1000], \(\gamma\): [0.0001, 0.001, $\dots$, 100]\};
DNN: \{activation: [\(relu\)], number of layers: [2, 3], layer width: [32, 64],
batch size: [\(64\)], learning rate: [\(0.001\)], \}.
We train the DNNs with the log-loss function for a total of 10 epochs using the ADAM optimizer.
To reduce variance due to random initialization, we train DNNs with 10 different random seeds 
and report the average of the runs.
We scale the input features using the maximum values from the training speakers for each test fold
before feeding the features into the classifiers.

For every test speaker, we run feature selection on the 
training speakers using likelihood ratio tests to determine whether 
individual features are significantly affected by mood. 
We retain features if the resulting \(p\)-value is less than \(0.05\).
We found that the total call duration was highly predictive of clinical outcomes.
As a result, we do not include it as a feature in our analysis to focus our study
on dialogue features that capture local interaction dynamics of the interviews.

\subsection{Results and Discussion}

We summarize the results for the classification tasks in Table~\ref{tab:clf_results}.
Consistent with previous work~\cite{gideon2016mood}, our results show that rhythm features are effective for detecting
both depression and mania. 
When using rhythm features alone, we obtain a maximum AUROC of 0.739 when predicting depression
using a logistic regression classifier and a maximum AUROC of 0.676 when predicting mania 
using a DNN classifier.
When using dialogue features alone, we obtain a maximum AUROC of 0.634 when predicting depression
using a logistic regression classifier and a maximum AUROC of 0.618 when predicting mania 
using a DNN classifier. 

Next, we study how combining the two feature sets, via early fusion,
affects the performance of predicting depression and mania.
When fusing rhythm and dialogue features, we obtain a maximum AUROC of 0.764 when predicting depression
using a SVM classifier and a maximum AUROC of 0.651 when predicting mania using a
DNN classifier. Augmenting rhythm features with dialogue features resulted in improved AUROCs for 
all three classifiers when detecting depression, suggesting that interaction patterns are 
complementary to speech rhythm patterns. This pattern was not true for detecting mania, however.
For detecting mania, we found that none of the classifiers was able to make use of the 
additional dialogue features to get improved AUROCs over using rhythm features alone.
This could be due to the fact that the relatively small number of manic episodes
in our dataset makes it hard for the trained classifiers to generalize to unseen test speakers.

\begin{table}[t]
\caption{Detecting mania/depression from clinical calls using three classifiers. 
Results shown are average AUROCs across all test speakers. 
Early fusion was used to combine rhythm and dialogue features.}
~\label{tab:clf_results}
\centering
\begin{tabular}{cccc}
\toprule
&  & \multicolumn{2}{c}{\textbf{AUROC}}  \\
\textbf{Classifier}          & \textbf{Features} & \textbf{Depressed} & \textbf{Manic} \\
\midrule
\multirow{3}{*}{LR}  & rhythm   & 0.739 & 0.658  \\
                     & dialogue & 0.634 & 0.606   \\
                     & both     & 0.761 & 0.641   \\
\midrule
\multirow{3}{*}{SVM} & rhythm   & 0.724 & 0.650   \\
                     & dialogue & 0.618 & 0.547   \\
                     & both     & 0.764 & 0.630   \\
\midrule
\multirow{3}{*}{DNN} & rhythm   & 0.729 & 0.676   \\
                     & dialogue & 0.617 & 0.618   \\
                     & both     & 0.761 & 0.651   \\
\bottomrule

\end{tabular}
\end{table}


\section{Conclusion}
In this work we showed that high-level dialogue features can be used to quantify 
interaction dynamics in clinical interviews, highlighting how changes in mood episodes 
can significantly affect the values of the features.
Additionally, we showed that dialogue features
can be used to augment prosodic features to improve automatic detection of
depression severity in clinical interviews.
For future work, we plan consider building gender-specific models for mood prediction,
as those have shown promise in emotion recognition tasks~\cite{zhang2016cross}.


  

\bibliographystyle{IEEEtran}

\bibliography{mybib}

\end{document}